\documentclass[twocolumn,secnumarabic,amssymb, nobibnotes, aps, prx]{revtex4-1}

\usepackage{amsmath}
\usepackage{mathtools}
\usepackage{subfigure}
\usepackage{subfig}
\usepackage{array,multirow}
\usepackage{booktabs}

\usepackage{algorithmic}
\usepackage{graphicx}
\usepackage{textcomp}
\usepackage{xcolor}
\usepackage{subfig}
\usepackage{caption}

\newcommand{\CrI}[1]{{$\mathrm{CrI_3}$ #1}}
\newcommand{\eg}[1]{{\textit{e.g.,} #1}}
\newcommand{\ie}[1]{{\textit{i.e.,} #1}}
\newcommand{\CrGeTe}[1]{{$\mathrm{CrGeTe_3}$#1}}
\newcommand{\CrBr}[1]{{$\mathrm{CrBr_3}$#1}}
\newcommand{\FeCl}[1]{{$\mathrm{FeCl_2}$#1}}
\newcommand{\R}[1]{{$R\bar{3}$#1}}
\newcommand{\C}[1]{{$C2/m$#1}}

\begin{document}
\title{Critical behavior of ferromagnets $\rm CrI_3$, $\rm CrBr_3$, $\rm CrGeTe_3$, and anti-ferromagnet $\rm FeCl_2$: a detailed first-principles study}
\author{Sabyasachi Tiwari$^{1,2,3}$, Maarten~L.~Van~de~Put$^{1}$, Bart~Sor\'ee$^{2}$$^{4}$$^{5}$ and William~G.~Vandenberghe$^{1}$}
\address{$^1$ Department of Materials Science and Engineering, The University of Texas at Dallas, 800 W Campbell Rd., Richardson, Texas 75080, USA.}
\address{$^2$ Imec, Kapeldreef 75, 3001 Heverlee, Belgium.}%
\address{$^3$Department of Material Engineering, KU Leuven, Kasteelpark Arenberg 10, 3001 Leuven, Belgium.}%
\address{$^4$Department of Electrical Engineering, KU Leuven, Kasteelpark Arenberg 10, 3001 Leuven, Belgium.}%
\address{$^5$Department of Physics, Universiteit Antwerpen, Groenenborgerlaan 171, 2020 Antwerp, Belgium.}

\begin{abstract}
We calculate the Curie temperature of layered ferromagnets, chromium tri-iodide ($\rm CrI_3$), chromium tri-bromide (\CrBr), chromium germanium tri-telluride (\CrGeTe), and the N\'eel temperature of a layered anti-ferromagnet iron di-chloride (\FeCl), using first-principles density functional theory calculations and Monte-Carlo simulations.
We develop a computational method to model the magnetic interactions in layered magnetic materials and calculate their critical temperature.
We provide a unified method to obtain the magnetic exchange parameters ($J$) for an effective Heisenberg Hamiltonian from first-principles, taking into account both the magnetic ansiotropy as well as the out-of-plane interactions.
We obtain the magnetic phase change behavior, in particular the critical temperature, from the susceptibility and the specific-heat, calculated using the three-dimensional Monte-Carlo (Metropolis) algorithm.
The calculated Curie temperatures for ferromagnetic materials ($\rm CrI_3$, \CrBr~and \CrGeTe), match well with experimental values.
We show that the interlayer interaction in bulk \CrI with \R~stacking is significantly stronger than the \C~stacking, in line with experimental observations.
We show that the strong interlayer interaction in \R~\CrI results in a competition between the in-plane and the out-of-plane magnetic easy axis.
Finally, we calculate the N\'eel temperature of \FeCl~to be $\rm 47\pm8\,K$, and show that the magnetic phase transition in \FeCl~occurs in two steps with a high-temperature intralayer ferromagnetic phase transition, and a low-temperature interlayer anti-ferromagnetic phase transition. 
\end{abstract}
\maketitle

\section{Introduction}

Two-dimensional (2D) magnetic materials~\cite{sample10,sample14,sample8} have attracted immense attention for their possible use in a plethora of spin-based applications, ranging from spintronics~\cite{sample21}, valleytronics~\cite{sample37}, magnetic memories~\cite{sample22} to topologically protected magnons~\cite{sample23}.
Recently, 2D magnetic crystals like $\rm CrI_3$~\cite{sample19,sample11,sample24}, $\rm CrGeTe_3$~\cite{sample20}, as well as doped 2D magnetic materials, \eg doped graphene ~\cite{sample28} and metal doped tansition-metal dichalcogenides (TMDs)~\cite{sample14,sample8,sample30,sample31,sample32,sample36} have been realized.

Layered magnetic materials open a plethora of opportunities for realizing novel magnetic devices~\cite{L_AFM,interlayer_app,layer_app}.
In layered magnetic materials, the strength of interlayer and the intralayer magnetic interaction remains significantly different~\cite{sample11,sample20,sample14, Reyntjens_2020}, opening the possibility to control their interlayer interaction electrically~\cite{layer_app}.
Moreover, in layered anti-ferromagnets, \eg~$\rm CrCl_3$, recent experiments have revealed a two-step phase transition, with a high temperature in-plane ferromagnetic (FM) phase, and a low-temperature out-of-plane anti-ferromagnetic phase~\cite{CrCl3}.
The phenomenon of the two-step phase transition in layered anti-ferromagnets is interesting from the point of view of physics, as well as for the application of layered anti-ferromagnets in realizing novel devices for memory applications~\cite{interlayer_app}.

Theoretical understanding of magnetism in 2D layered materials is of great importance for their possible use in futuristic spin-based technologies~\cite{L_AFM,interlayer}.
Reliable quantification of the critical parameters using first-principles calculations, such as the critical temperature (Curie/N\'eel), is necessary for evaluating the candidacy of layered magnetic materials for their possible application~\cite{spintronics, S_problem}, and for designing newer layered materials and their heterostructures~\cite{Mat_design,interlayer_app}.
Although many 2D magnetic materials have recently been investigated, theoretical efforts in modeling the magnetic structure, and the calculation of critical temperatures have mostly remained confined to their monolayers ~\cite{approach_1,sample26,beyond_J}, ignoring
their layered bulk forms, which are more interesting in terms of applications~\cite{L_AFM} and are physically more stable~\cite{sample19,sample11,sample24,sample20}.
Several methods including, the mean-field approximation~\cite{CrI_bad,CrI_bad2}, the Ising model~\cite{beyond_J}, the random-phase approximation (RPA)~\cite{joren}, and the linear spin-wave~\cite{S_problem,sample26}, have been used to calculate the critical temperature and the magnetic phase transition of the monolayers of 2D magnetic materials.
However, a similar theoretical effort for modeling layered magnetic materials is missing.

Unfortunately, the magnetic structure of layered magnetic materials mostly has been modeled qualitatively~\cite{CrI_bilayer1,CrI_bilayer2, CrI_bilayer3, CrI_bilayer4,multilayer-2,multilayer-3}, not quantitatively~\cite{J1_J2,MFT2}.
Most theoretical works on layered magnetic materials have either focussed on explaining the experimental observations qualitatively, \eg in bilayer $\rm CrI_3$~\cite{CrI_bilayer1,CrI_bilayer2, CrI_bilayer3, CrI_bilayer4}, or have used inputs directly from experiments to calculate experimental observables, \eg~the Curie temperature of \CrGeTe~calculated in ref.~\cite{sample20}.
To our knowledge, the magnetic structure of layered materials has not been studied entirely from the first principles while accounting for full magnetic anisotropy. 
A study of the magnetic phase transition and critical temperature in bulk layered materials, which feature different in-plane and out-of-plane exchange interactions, is missing.

In this paper, we calculate the Curie temperature of bulk ferromagnets $\rm CrI_3$, \CrBr, \CrGeTe,~and study the magnetic phase transition of the layered anti-ferromagnet \FeCl~along with calculating its N\'eel temperature.
We model the magnetic interactions using first-principles density functional theory (DFT) calculations and study the magnetic phase transition using three-dimensional (3D) Monte-Carlo simulations.
In Section~\ref{s:method}, we introduce our model for the magnetic structure.
In Section~\ref{s:Curie_temp}, we investigate the difference in the magnetic interaction between \C~ and \R~stacked $\rm CrI_3$, and calculate the Curie temperatures of three ferromagnetic layered materials, $\rm CrI_3$, \CrBr, and \CrGeTe{}.
We obtain Curie temperatures of 72 K, 49 K, and 103 K for $\rm CrI_3$, \CrBr, and \CrGeTe, respectively, which we show to be in agreement with their experimentally measured Curie temperatures.
Further, we calculate the N\'eel temperature of a layered anti-ferromagnet, iron di-chloride (\FeCl), and find its N\'eel temperature to be $47\pm8$ K, and show that \FeCl~undergoes a double phase transition.
In Section IV, we present our computational model detailing how we obtain the magnetic interactions from the first-principles DFT calculations and estimate the critical temperature using three-dimensional (3D) Monte-Carlo simulations.
In section V, we conclude.

\section{Magnetic structure model}{\label{s:method}}

We start from the general form of the Heisenberg Hamiltonian,
\begin{equation}
    H
    = -\sum_{i\neq j} {\mathbf{S}}_{i}J_{ij}{\mathbf{S}}_{j}-D\sum_{i}({S}_{i}^z)^2.
    \label{e:hamiltonian}
\end{equation}
The first term is the exchange term between the $i^{\rm th}$ and the $j^{\rm th}$ magnetic atom where ${\mathbf{S}_i}$ are the spin-operator for the $i$th magnetic atom.
$J_{ij}$ measures the strength of the exchange interaction between the $i^{\rm th}$ and the $j^{\rm th}$ magnetic atom.
We treat the Heisenberg Hamiltonian in the classical approximation where the spin-operator is the local magnetic moment (magnetization) associated with the magnetic atom.
The second term is the onsite anisotropy term, with $D$ being the strength of the onsite anisotropy.
The magnetic moments ${\mathbf{S}}$ in Eq.~(\ref{e:hamiltonian}) are vectors with ${\mathbf{S}}=S^x\mathbf{x}+S^y\mathbf{y}+S^z\mathbf{z}$.
$J_{ij}$ is a tensor,

\begin{equation}
J_{ij}=
\begin{bmatrix}
J_{ij}^{xx} & J_{ij}^{xy} & J_{ij}^{xz} \\
J_{ij}^{yx} & J_{ij}^{yy} & J_{ij}^{yz} \\ 
J_{ij}^{zx} & J_{ij}^{zy} & J_{ij}^{zz}
\end{bmatrix}.
\label{e:tensor}
\end{equation}

\begin{figure}[ht]
	\centering
   \includegraphics[width=0.9\columnwidth]{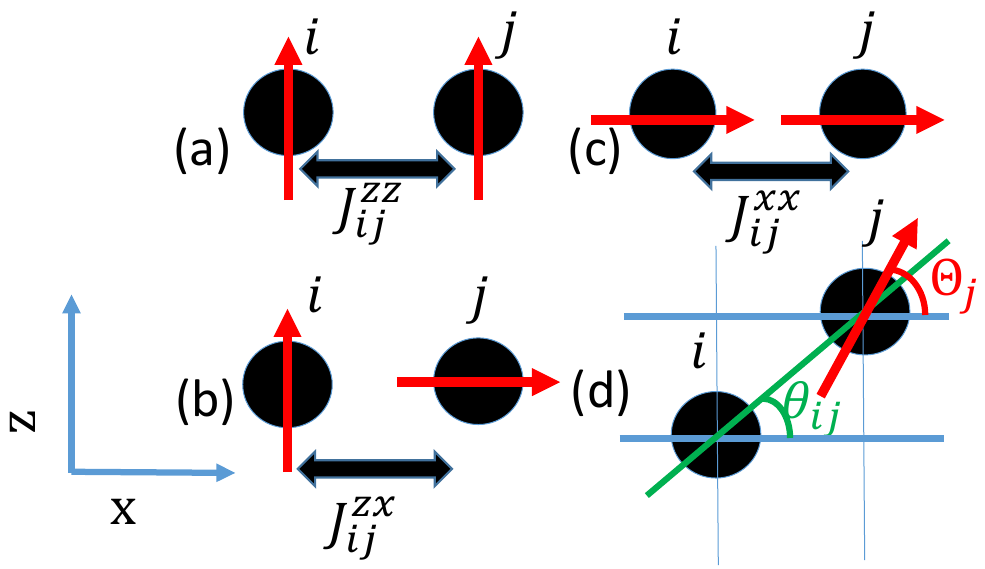}
    \caption{Illustration of the elements of matrix $J_{ij}$ for various magnetic axis orientation in different directions for atoms $i$ and $j$ (a), (b) and (c), and the azimuthal angle $\theta_{ij}$ between atoms $i$ and $j$ in two different layers of a magnetic material and the magnetic anisotropic angle $\Theta_j$ for the magnetic atom $j$ (d).}
	\label{f:illustration2}
\end{figure}

We illustrate various elements of the $J_{ij}$ tensor in Fig.~\ref{f:illustration2}. 
Figures~\ref{f:illustration2} (a) and (c) show that $J_{ij}^{xx}$ and $J_{ij}^{zz}$ are the coupling strength when the magnetic axis for both the atoms $i$ and $j$ are both oriented in the $\mathbf{x}$ and $\mathbf{z}$ direction, respectively, Fig.~\ref{f:illustration2} (b) shows that $J_{ij}^{zx}$ is the coupling strength when the magnetic axis of atom $i$ is oriented in the $\mathbf{z}$ direction and the magnetic axis of atom $j$ is oriented in the $\mathbf{x}$ direction.
It is important to understand that an atom at site $i$ and $j$ have a difference in their magnetic axis measured by a ``magnetic angle ($\Theta$)" which is completely different than the ``geometric angle ($\theta$)" (illustrated in Fig. 1d) measuring the angle between the axis connecting atoms $i$ and $j$ in the plane.

The diagonal elements of $J_{ij}$ represent the collinear exchange interaction, \ie~the exchange interaction between the same components of the magnetic moments of atoms $i$ and $j$.
The off-diagonal elements of the tensor $J_{ij}$ represent a non-collinear exchange interaction, \ie the exchange interaction between different components of the magnetic moments of atoms $i$ and $j$.

We model the pair-wise exchange interaction tensors $J_{ij}$ as a parameterized continuous function of distance ($r_{ij}=r_i-r_j$),  azimuthal angle ($\theta_{ij}$, illustrated in Fig.~\ref{f:illustration2}(d)), and the in-plane angle $\phi_{ij}$, between atoms $i$ and $j$.
Typically for monolayers, only up to the nearest-neighbor exchange interaction has been used by most of the previous works~\cite{sample26,CrI_bad,CrGeTe_bad} because going beyond the nearest-neighbor interaction increases the number of non-zero $J_{ij}$ tensors rapidly~\cite{beyond_J}.
We go beyond the nearest-neighbor interaction and also take into account the out-of-plane exchange interactions for layered materials, but a continuous function to approximate $J_{ij}$ helps in reducing the number of $J_{ij}$ tensors.
% leading to a greatly increased number of $J_{ij}$ tensors.
The elements of the $J_{ij}$ tensor then read, $J^{\alpha\beta}(r_{ij},\theta_{ij},\phi_{ij})$ where, $\alpha,\beta \in$~\{$x,y,z$\}.
Using the parametrized $J^{\alpha\beta}(r_{ij},\theta_{ij},\phi_{ij})$ yields a parametrized Heiseberg Hamiltonian.
Since we only consider layered materials in this paper with a 3-fold rotational symmetry axis, we assume the exchange interaction to be isotropic in-plane and to be independent of the in-plane angle.
We further denote $J(r,\theta)$ omitting the in-plane angle $\phi_{ij}$.

\begin{figure*}[ht]
   \centering
   \includegraphics[scale=0.5]{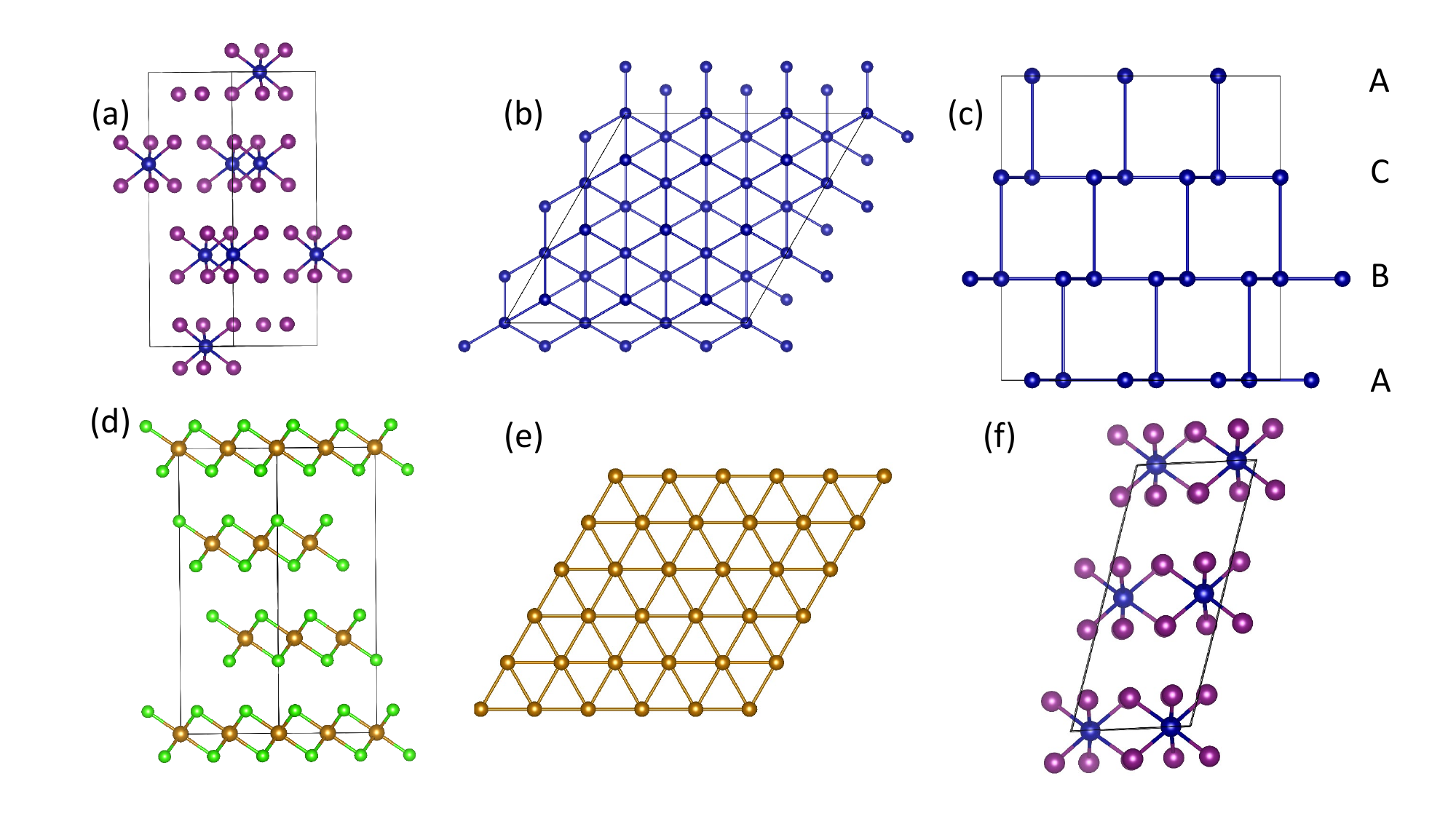}
   \caption{(a) Side view of \R~\CrI. (b) Top view of the net of Cr atoms in \R~$\rm CrI_3$. (c) Side view of the Cr atoms in \R~\CrI with A, B, and C stacks are shown explicitly. (d) Side view of \FeCl~bulk. (e) Top view of the net of Fe atoms in \FeCl~bulk. (f) Side view of \C~$\rm CrI_3$.}
  \label{f:CrI_R3}
\end{figure*}

The functions $J^{\alpha\beta}(r_{ij},\theta_{ij})$ are obtained by fitting the energies obtained from the parametrized Heisenberg Hamiltonian to the total energies calculated using the first-principles calculations for several magnetic configurations including, ferromagnetic, ferrimagnetic, and anti-ferromagnetic configurations.
Having obtained optimal $J^{\alpha\beta}(r_{ij},\theta_{ij})$, we calculate the critical temperature by simulating the phase change of the parameterized Heisenberg Hamiltonian of Eq.~(\ref{e:hamiltonian}), using the 3D Monte-Carlo algorithm.
The details on our computational procedure including the exact parametrized functional form, the algorithm to obtain the parameters of the functional form, and the calculation of critical temperatures using MC simulations, are provided in Section~\ref{s:method2}.

Our method of parameterizing Heisenberg Hamiltonian only relies on ground state calculations of supercells and differs from the computationally more expensive spin-spiral method which relies on the generalized Bloch condition~\cite{ref1,ref2}. 
Fitting a functional form for the $J$-parameters, decaying at long distances, is computationally more efficient than the spin-spiral method and can be expanded to the study of lattices with random magnetic dopants.

\section{Results}{\label{s:Curie_temp}}

We first calculate the Curie temperature of layered ferromagnets $\rm CrI_3$, \CrBr, and \CrGeTe, and compare with experimental data, revealing a much-improved match compared to previous calculations.
We then apply our method on the layered anti-ferromagnet \FeCl,~and predict its N\'eel temperature and a two-step phase transition behavior.

\subsection{Crystal structure of $\rm CrI_3$, \CrBr, \CrGeTe~and \FeCl{}}

Figure~\ref{f:CrI_R3} (a-c) show the crystal structure of \CrI with the \R~space group.
The in-plane lattice of chromium atoms shown in Fig.~\ref{f:CrI_R3} (b) has a hexagonal lattice structure for all the three mentioned materials.
The \R~structure (Fig.~\ref{f:CrI_R3}) (c) resembles the ABC stacking of hexagonal materials, with the unit-cell comprising of six Cr atoms.
Figure~\ref{f:CrI_R3} (d) and (e) show the lattice structure of $\rm FeCl_2$.
In $\rm FeCl_2$, each Fe atom has six nearest-neighbors (NN) and twelve next-nearest-neighbors (NNN) in-plane, as shown in Fig.~\ref{f:CrI_R3} (e).
In the out-of-plane direction, there are six neighboring (O-NN) Fe atoms.
Figure~\ref{f:CrI_R3} (f) shows the side-view of the lattice structure of \C~\CrI.
The in-plane lattice structure is similar to that of \R~\CrI but the out-of-plane stacking is different, and the Bravais lattice is monoclinic in nature instead of rhombohedral as in the case of \R~\CrI.

All three chromium compounds and \FeCl~have a low-temperature stacking order commensurate with the \R~space group.
Bulk \CrI and \CrBr~exhibit a structural phase transition from the \C~to the \R~space group at low temperature (below 200 K).
The crystal structures of the magnetic compounds shown in Fig.~\ref{f:CrI_R3}, agree well with their experimental lattice structures~\cite{sample11, sample20,FeCl2_exp}.
More details on structural parameters are provided in the supplementary document.

\begin{figure*}[t]
  \subfigure[]{\includegraphics[width=\columnwidth]{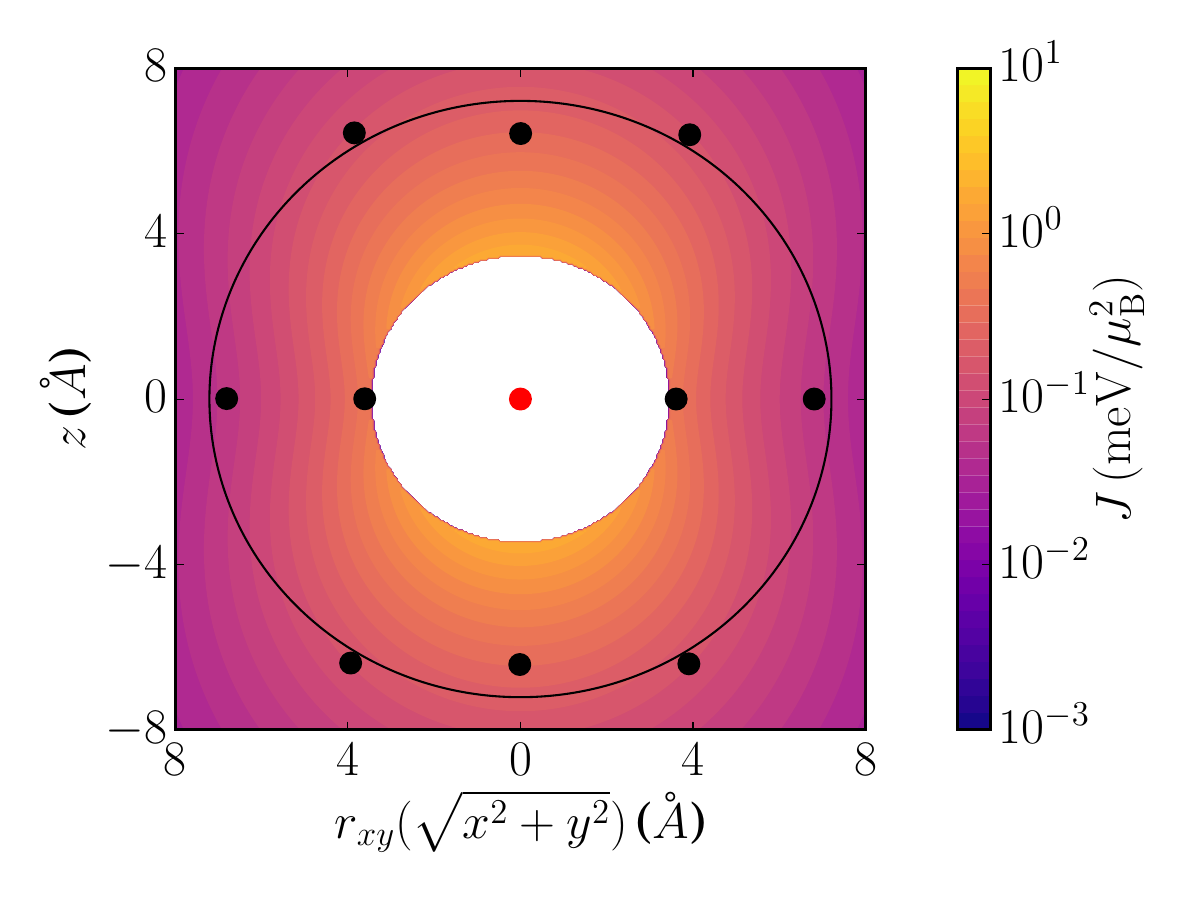}}\quad   
 \subfigure[]{\includegraphics[width=\columnwidth]{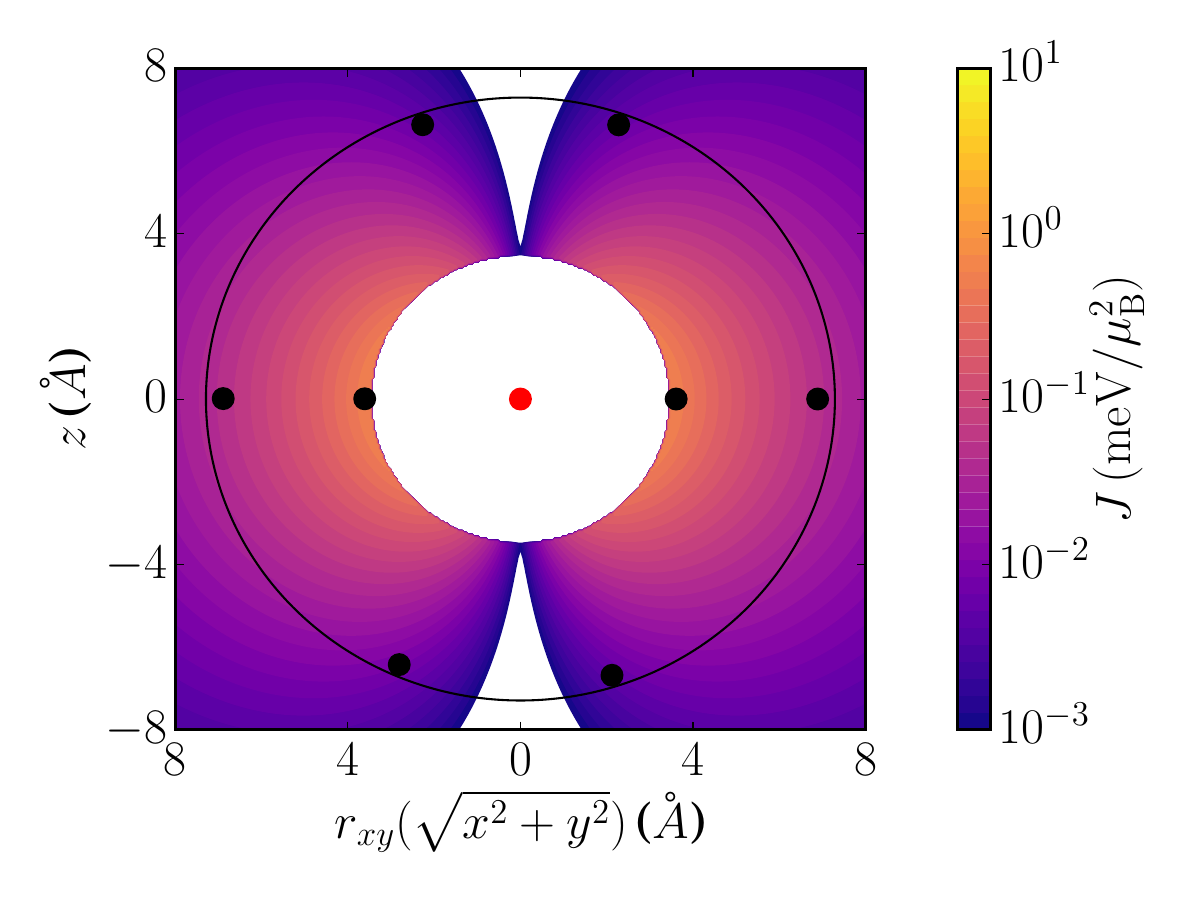}}
  \caption{(a)Shows the $J$ parameter ($J^{zz}(r,\theta)$) when the magnetic easy axis is out-of-plane for (a)\R~\CrI and (b)\C~\CrI. The circle shows the range of interaction ($r_{\rm c}=6.81$ \AA~for \R~and $r_{\rm c}=7.01$ \AA~for \C) with respect to the atom at the centre (in red). The atoms inside the interaction circle are the ones considered while obtaining the $J$ parameters.}
 \label{f:J_R3}
\end{figure*}
\subsection{Comparing the magnetic interaction in \R~\CrI~and \C~\CrI{}}

In this section, we analyze \CrI in its \R~phase which is the most stable phase as well as its \C~phase, which is stable at high temperature. 
However, the high-temperature \C~phase is of interest because \CrI bi-layer or tri-layered \CrI have been experimentally shown to exist both in the \R~and the \C~phase~\cite{sample24}, and stacking in bilayer \C~or \R~\CrI has been of much theoretical interest~\cite{CrI_bilayer1,CrI_bilayer2, CrI_bilayer3, CrI_bilayer4}.
%In this section, we compare the magnetic interaction in \R~and the \C~phases of $\rm CrI_3$.

\subsubsection{Difference between the exchange interaction in \R~and \C~\CrI{}}\label{S:exchange}

Figure~\ref{f:J_R3} shows the exchange interaction strength between the $z$ components of two magnetic moments $J^{zz}(r,\theta)$ for \CrI with (a) \R~stacking and (b) \C~stacking.
Here, the interaction is between the central Cr atom and the surrounding atoms, within a radius of interaction ($r_{\rm c}$=6.81 \AA~for \R~and $r_{\rm c}$=7.01 \AA~for \C).
The color scale indicates the interaction strength $J^{zz}(r,\theta)$ with respect to the central Cr atom.

We find that the in-plane interaction strength in both the \C~\CrI and the \R~\CrI is similar but the out-of-plane interaction is significantly weaker in \C~\CrI compared to \R~$\rm CrI_3$.
This result is in line with the experimental observation where the out-of-plane interaction was found to be comparable to the in-plane interaction in \R~$\rm CrI_3$~\cite{sample11}.
The weak ferromagnetic out-of-plane interaction in a few-layer \C~phase of \CrI has also been reported experimentally~\cite{sample24}.
Our result is also consistent with previous theoretical works on \R~and \C~stacked bilayer~\CrI in which, the difference in the out-of-plane interaction in \C~and \R~\CrI was attributed to the difference in their out-of-plane super-superexchange interaction~\cite{super-superexchange}.

Referring to Fig.~\ref{f:Curie}, we find that the critical exponent $\beta$ is lower for the \C~phase of \CrI and the reason for that is the low out-of-plane interaction strength in \C~\CrI as shown in Fig.~\ref{f:J_R3}.
The weak out-of-plane interaction makes the out-of-plane magnetic orientation unstable at temperatures even below the Curie temperature in \C~$\rm CrI_3$, resulting in a lower $\beta$.
\subsubsection{Curie temperature and critical behaviour}
\begin{figure}[ht]
	\centering
    \includegraphics[width=\columnwidth]{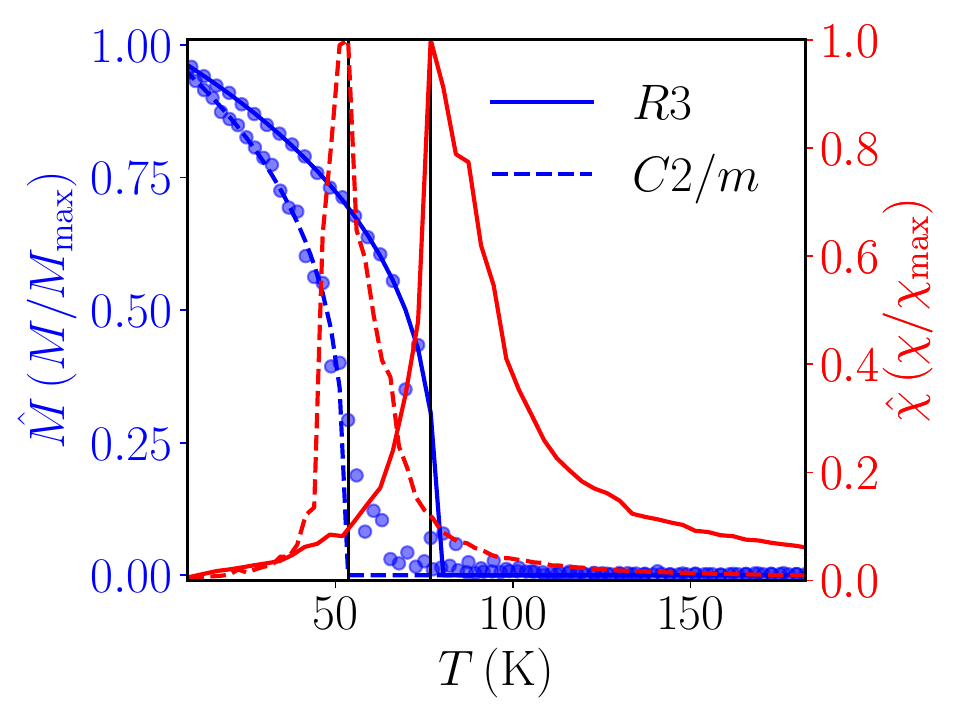}
    \caption{Shows the magnetization Vs temperature (left) and susceptibility vs temperature (right) for \CrI with \R~space-group (blue) and \C~space-group (red). Solid lines show the Curie-Weiss function ($M\propto(T-T_{\mathrm{c}})^{\beta}$) fit, dots show the magnetization ($M$) obtained from the MC simulation, dotted lines show the susceptibility obtained from the MC simulation.}
	\label{f:Curie}
\end{figure}

Figure~\ref{f:Curie} shows the magnetization ($M$) and the susceptibility ($\chi$) as a function of temperature for the \R~and the \C~phases of $\rm CrI_3$.
We fit the Curie-Weiss function ($M\sim(T_{\mathrm{c}}-T)^\beta$) to the magnetic moment obtained from the MC simulations for both the \R~and \C~$\rm CrI3$.
From our fit, shown in Fig.~\ref{f:Curie}, we find $\beta=0.312$ for \R, while for \C, we find a slightly lower value of $\beta=0.28$.

The susceptibility peaks at around 72 K for the \R~and 55 K for the \C~phase of $\rm CrI_3$.
The temperature at which the susceptibility peaks is the Curie temperature.
Therefore, we find that the stacking does not make a significantly big difference ($< 20$) K in the Curie temperature of bulk $\rm CrI_3$.

\subsubsection{Interplay between geometric and magnetic anisotropy in \CrI{}}\label{S:exchange}
Up to now, we discussed the geometric anisotropy of \C~and \R~$\rm CrI_3$.
In this section, we focus on the magnetic anisotropy of \C~and \R~phases of $\rm CrI_3$.
\begin{figure}[ht]
   \centering
  \includegraphics[scale=0.42]{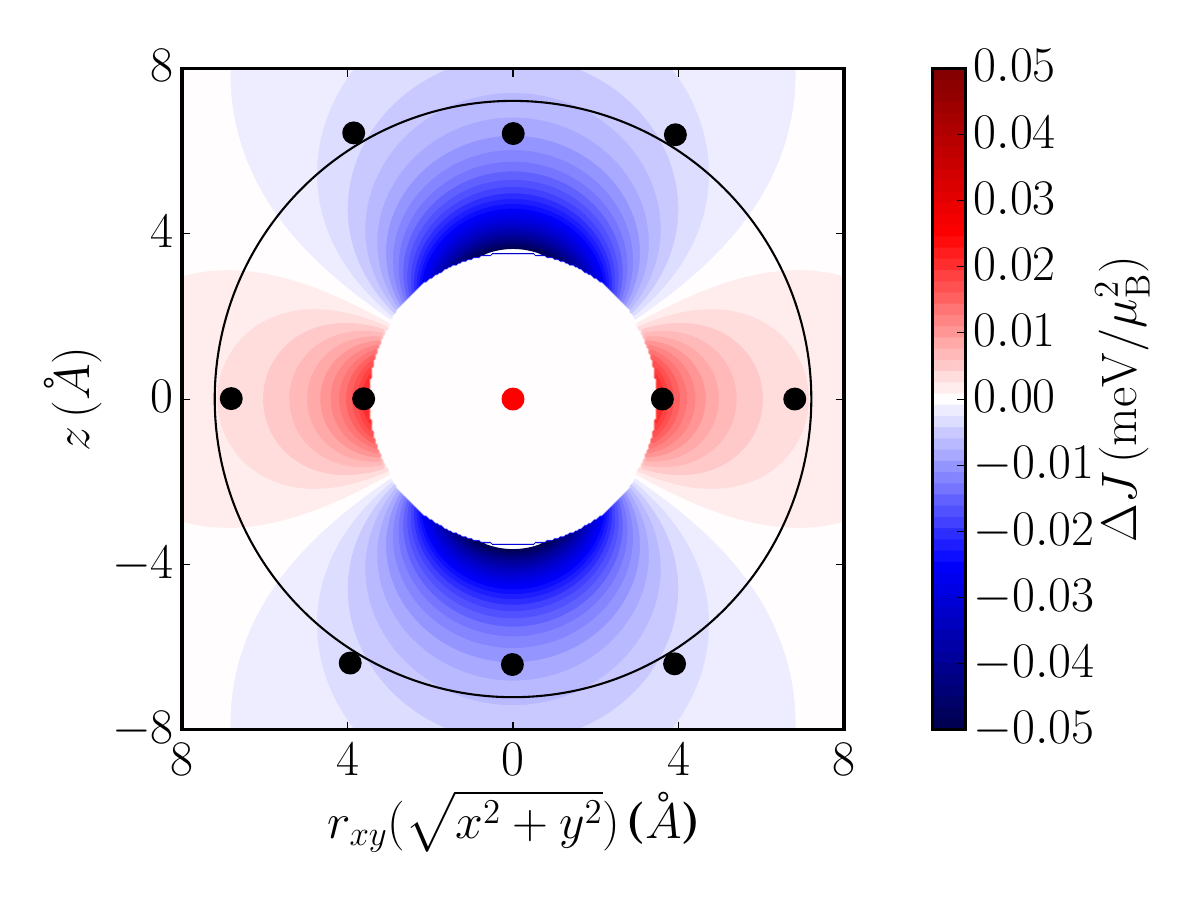}\quad
   \caption{The interplay between the magnetic and geometric anisotropy. The difference in the $J$ parameter ($\Delta J=J^{zz}(r,\theta)-J^\parallel(r,\theta)$) for \CrI bulk when the magnetic easy axis is out-of-plane ($001$) Vs when the magnetic easy axis lies in-plane ($010$)}
 \label{f:J_R3_LS}
\end{figure}

Figure~\ref{f:J_R3_LS} shows the difference between the $J$ parameters ($\Delta J=J^{zz}(r,\theta)-J^\parallel(r,\theta)$, where $J^\parallel(r,\theta)=J^{xx}(r,\theta)=J^{yy}(r,\theta)$) when the magnetic axis is oriented out-of-plane ($z$) and when the magnetic axis is oriented in-plane ($\parallel$).
The difference in the out-of-plane interaction is negative, and the difference in the in-plane interaction is positive.
This means that the out-of-plane magnetic axis maximizes the intralayer interaction, and a magnetic axis along the in-plane direction maximizes the interlayer interaction.
Hence, we see a competition between the magnetic orientation (along the easy axis or not), the intralayer interaction, and the interlayer interaction in \R~$\rm CrI_3$.

The out-of-plane magnetic axis is the easy-axis in \R~\CrI because, in the in-plane direction of \R~ $\rm CrI_3$, there are three nearest-neighbors, whereas, in the out-of-plane direction, there are only two nearest-neighbors.
Therefore, an out-of-plane alignment of the magnetic moments minimizes the total energy gain because it maximizes the intralayer interaction in \R~$\rm CrI_3$.

We do not find any competition between geometric and magnetic anisotropy in \C~$\rm CrI_3$.
An out-of-plane magnetic axis maximizes the intralayer interaction but the interlayer interaction is insensitive to the change in the magnetic axis in \C~$\rm CrI_3$, and its easy-axis turns out to be in the out-of-plane direction.
The absence of competition between magnetic and geometric anisotropy in \C~\CrI, which has a weak out-of-plane super-superexchange interaction~\cite{CrI_bilayer2,CrI_bilayer3,CrI_bilayer4}, implies that the interplay between geometric and magnetic anisotropy in \R~\CrI is a result of strong out-of-plane super-superexchange interaction.
\subsection{The Curie temperature of $\rm CrI_3$, \CrBr~and \CrGeTe{}}

\begin{table}[ht]
\caption{Curie temperatures for \CrI~\CrBr~and~\CrGeTe~} 
\centering
\resizebox{\columnwidth}{!}{
\begin{tabular}{c c c c} % centered columns (4 columns)
\hline%inserts double horizontal lines
\addlinespace[1mm]
Compound  &  \CrI (\R){} & \CrBr~(\R){} & \CrGeTe~(\R) \\ [0.5ex]
\hline
\addlinespace[1mm]
$T_{\rm c}$ (our work, bulk) & $72$ K & $49$ K& $103$ K  \\
$T_{\rm c}$ (our work, single-layer (SL)) & $69$ K & $39$ K& $65$ K  \\
$T_{\rm c}$ (experimental, bulk) & $61$-$70$ K\cite{sample11,sample24,crystal} & $37$-$47$ K\cite{Cr_Br_exp,crystal}& $66$-$75$ K\cite{sample20}  \\
$T_{\rm c}$ (experimental, SL) & $45$ K\cite{sample24} & $34$ K\cite{CrBr_mon} & $45$ K\cite{sample20}  \\
$T_{\rm c}$ (theory, $S>1$) & $161$ K\cite{sample10}, $95$ K\cite{CrI_bad2} & $  $ & $314$ K\cite{sample10} \\
$T_{\rm c}$ (theory ($S=1$), SL) & $46$ K\cite{CrI_bad} & $41$ K\cite{CrI_bad} & $57$ K\cite{CrGeTe_bad}, $130$ K\cite{CrGeTe_bad_2} \\
\hline 
\end{tabular}
}
\label{t:Curie} 
\end{table}

Table~\ref{t:Curie} shows the calculated Curie temperature and a comparison with experimental and previous mono/single-layer (SL) calculations~\footnote{the parameters of the functional-form and the saturation magnetization is provided in supplimentarty tables S1-S5}.
The previous theoretical calculations, which used the mean-field approximation and ignored magnetic anisotropy, labeled as ``$T_{\rm c}$ (theory, SL)", are off by $318$\% for the SL \CrGeTe~and 163\% for $\rm CrI_3$, from their experimental values, respectively.
Whereas, in our calculation, the relative deviation from the experimental value for bulk compounds is as low as 37\%, 13\%, and 3\% for bulk \CrGeTe, \CrBr, and $\rm CrI_3$, respectively.
For monolayers in our calculations, the relative deviation from the experimental value is, 53.3 \%, 14.7 \%, and 44.4 \% for \CrI, \CrBr, and \CrGeTe, respectively.

Another set of calculations labeled as ``$T_{\rm c}$ (theory ($S=1$), SL)", also used the mean-field approximation but used $T_c=\frac{3J}{(2k_B )}$ instead of the more generally used $T_c=\frac{3J}{(2k_B )} S^2$.
In those mean-field calculations ($T_{\rm c}$(theory($S=1$),SL)), the relative deviation from the experimental value for monolayers was as low as 26.6\%, 20\%, and 2\% for \CrGeTe, \CrBr, and $\rm CrI_3$, respectively.
In the $(S=1)$ calculations, it appears that not accounting for $S^2$ compensated for the usual overestimation observed when using the mean-field equation to predict the Curie temperature of 2D magnets~\cite{joren2}. 
However, ignoring $S^2$ to compensate for the overestimation using mean-field cannot be expected to be a reliable strategy when studying materials with a significantly larger or smaller magnetic moment. 
On the other hand, our method does account for magnetic moment and provides a relatively good agreement to the experimental Curie temperatures for most of the ferromagnetic compounds.

We also calculate the critical exponent $\beta$ of the Curie-Weiss function ($M\sim (T_{\mathrm{c}}-T)^\beta$), which determines the near-criticality behavior of the magnetic materials.
We obtain a $\beta=0.312$ for $\rm CrI_3$, and $\beta=0.28$ and $\beta=0.341$ for \CrBr~and \CrGeTe, respectively.
Recently, experimental work in Ref.~\cite{critical} has reported the value of $\beta$ for \R~\CrI to be $0.32$.
Just like the Curie temperature, the critical exponent $\beta$ we obtain for \CrI is very close to the experimentally observed $\beta$ for $\rm CrI_3$.

\begin{figure}
   \centering
  \includegraphics[width=\columnwidth]{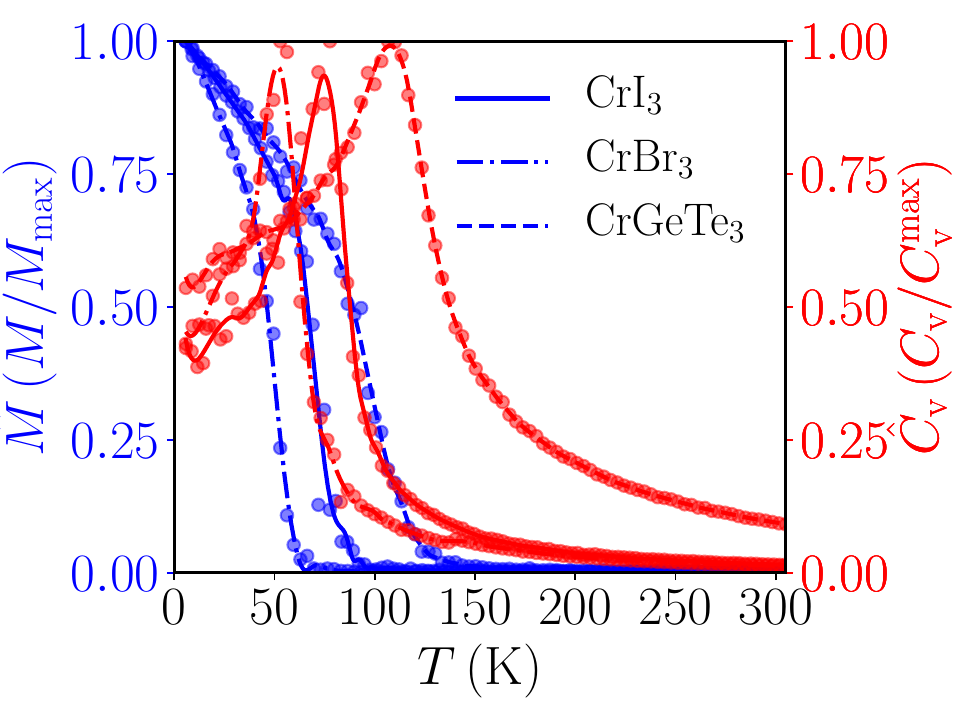}
  \caption{The magnetization vs temperature (blue) and the specific-heat vs temperature (red), of bulk \CrBr,\CrGeTe,$\rm CrI_3$. Solid lines show the interpolated data using a savgol filter while dots show the data obtained from the MC simulations. The saturation magnetization for \CrBr,\CrGeTe, and $\rm CrI_3$ are 2.7 $\rm \mu_B$, 2.7 $\rm \mu_B$, and 2.9 $\rm \mu_B$, respectively.}
 \label{f:Curie_Cr}
\end{figure}

Figure~\ref{f:Curie_Cr} shows the average magnetization ($\hat{M}$), and the specific-heat ($\hat{C}_{\rm v}$) as a function of temperature ($T$) for \CrBr, \CrGeTe,~and $\rm CrI_3$.
All three compounds are ferromagnets and show a transition from the paramagnetic phase to a ferromagnetic phase at the onset of the Curie temperature, which is the temperature at which the specific-heat peaks.
However, below the Curie temperature, we observe a small fluctuation in the specific-heat even though the magnetization remains stable.
We find that the reason for such small fluctuations in $\hat{C}_{\rm v}$ is the difference in the strength between the in-plane and the out-of-plane exchange interaction, which we refer as geometric anisotropy, as well as magnetic anisotropy.

\begin{figure}[ht]
	\centering
    \includegraphics[width=\columnwidth]{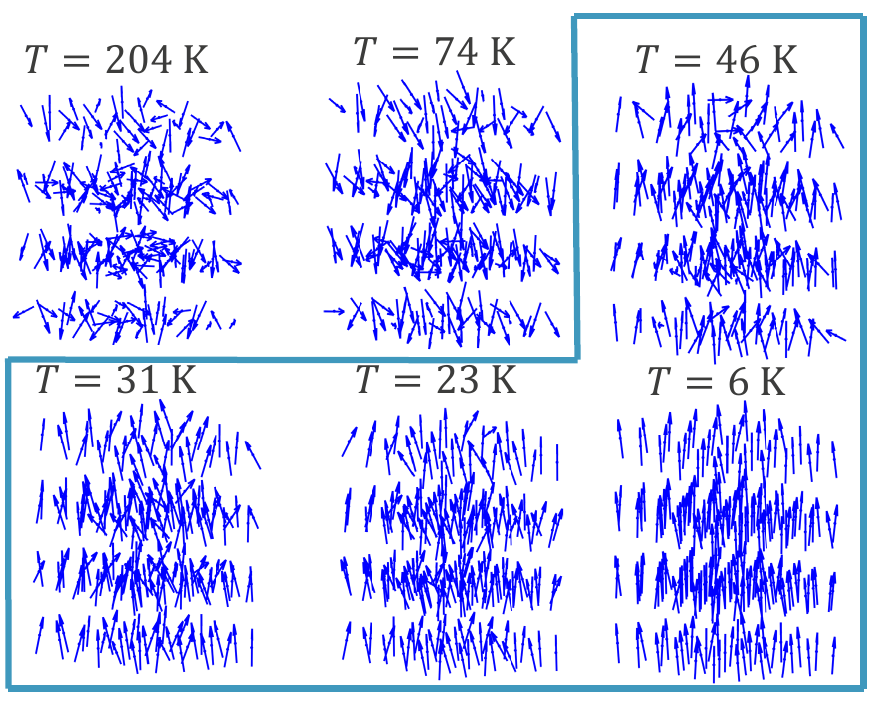}
    \caption{The magnetic orientation of a supercell of \R~\CrI at various temperatures. Magnetic orientations within the demarcated box are the orientations below the Curie temperature.}
	\label{f:Curie_or}
\end{figure} 

To visualize the consequence of geometric and magnetic anisotropy on the magnetic order, we depict the ground state magnetic configuration of \R~\CrI at various temperatures in Fig.~\ref{f:Curie_or}.
At a temperature of $\rm 204\, K$, the magnetic moments orient randomly, which implies a paramagnetic phase.
At 74 K, which is slightly above the Curie temperature, we see that ferromagnetic domains start forming within layers but interlayer magnetic axis orientation remains randomized, resulting in a low average magnetization.
At 46 K, which is below the Curie temperature of \R~$\rm CrI_3$, we see a short-range magnetic order within the layers but a preferred direction of magnetic orientation is missing.
We see some layers with an in-plane magnetic axis and some layers with an out-of-plane magnetic axis, which is a direct consequence of the geometric and magnetic anisotropy, hindering a preferred magnetic axis orientation.
At $\rm 31\,K$, and below, we see the ferromagnetic phase with the magnetic axis starting to orient in the out-of-plane direction with some perturbations which last till 23 K.
Finally, at 6 K, we see a perfectly aligned ferromagnetic phase with a magnetic axis in the out-of-plane direction.

\subsection{Double phase transition in \FeCl{}}

\begin{figure}[ht]
   \centering
  \includegraphics[width=\columnwidth]{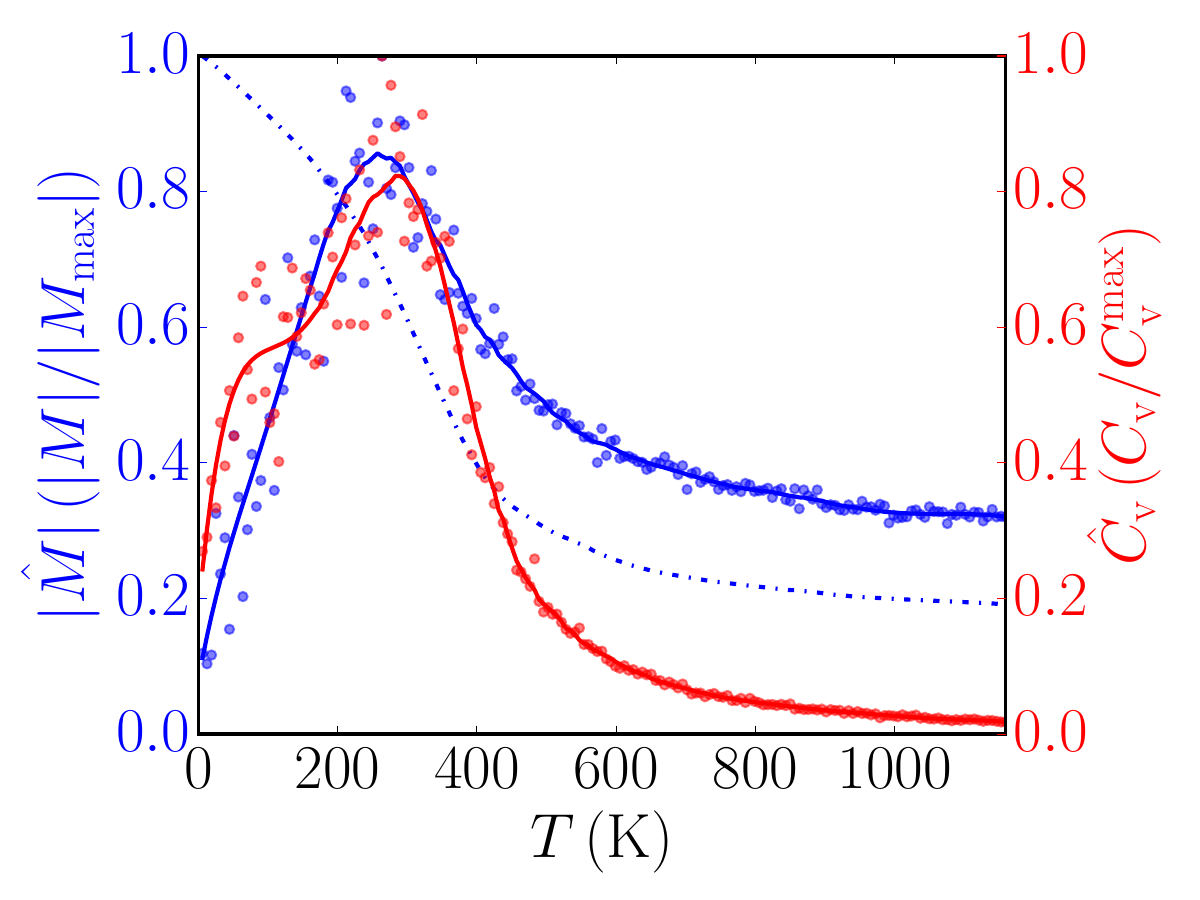}
  \caption{Solid lines show the interpolated magnetization vs temperature (blue) and the specific-heat vs temperature (red), of bulk \FeCl~obtained from the MC simulations. Dots show the data obtained from the MC simulations. The dotted line shows the magnetization vs temperature for one of the layers of \FeCl.  The maximum magnetizations are $|M_{\rm max}|\approx0.9\,\mu_B$ at 250 K, and $|M_{\rm sat}|\approx0.1\, \mu_B$ at 6 K, respectively.}
 \label{f:Curie_FeCl}
\end{figure}
 
Figure~\ref{f:Curie_FeCl} shows the absolute value of the average magnetization ($\hat{|M|}$) (blue dots) and the specific-heat ($\hat{C}_{\rm v}$) (red dots) as a function of temperature, obtained from the MC simulations for bulk \FeCl.
The solid lines in Fig.~\ref{f:Curie_FeCl} show the interpolated specific-heat (red line) and the interpolated magnetization (blue line) as a function of temperature.
The dotted line shows the average magnetization as a function of temperature for a single layer of bulk \FeCl.
Unlike in Fig.~\ref{f:Curie_Cr} where, we have plotted the average magnetization ($\hat{M}$), in Fig.~\ref{f:Curie_FeCl}, we have plotted the absolute value of the magnetization ($\hat{|M|}$) in order to separate the paramagnetic phase from the anti-ferromagnetic phase (for further details, see Eq.~(\ref{e:magnetization}) and our discussion thereof).
The paramagnetic phase at high temperature has higher absolute magnetization than the anti-ferromagnetic phase at low temperature because the magnetic moments orient themselves randomly in each MC step, whereas, for the anti-ferromagnetic phase the magnetic moments orient themselves in an opposite direction, causing the absolute magnetization to go to zero.

\begin{figure}[ht]
   \centering
  \includegraphics[width=\columnwidth]{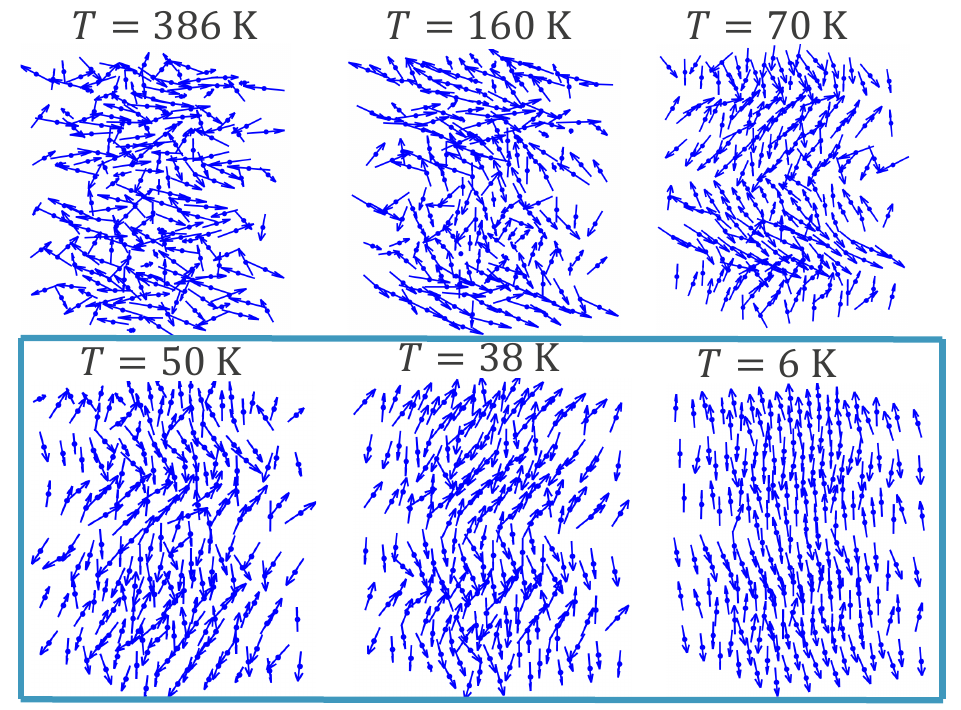}
   \caption{Shows the magnetic ground state of a $5\times 5\times 2$ supercell of \FeCl~at various temperatures obtained from the MC simulations. Magnetic orientations within the demarcated box are the orientations below the N\'eel temperature. }
  \label{f:FeCl2_GS}
\end{figure}

The specific-heat plot in Fig.~\ref{f:Curie_FeCl} shows one peak and one cusp, which implies that \FeCl~undergoes two magnetic phase transitions.
The high-temperature phase transition, which is the most prominent phase transition, occurs near a temperature close to 250 K.
At this temperature, we observe from the magnetization of a single layer \FeCl (dotted line in Fig.~\ref{f:Curie_FeCl}) that the layers of \FeCl~undergo a ferromagnetic phase transition.
However, the magnetization per Fe atom in the entire sample of \FeCl~reaches a maximum value of $0.9\,\mu_{\rm B}$, which is less than its maximum value of $3.5 \mu_{\rm B}$.
This low magnetization in bulk \FeCl~after the high-temperature phase transition is observed because even though each of the layers undergo a ferromagnetic phase transition, their alignment in the out-of-plane direction remains paramagnetic, reducing the overall magnetization of the bulk sample.

Lowering the temperature below 250 K, a second magnetic phase transition occurs between a temperature range of $40-55$ K.
This low-temperature phase transition is unique because, instead of showing a peak, the interpolated specific-heat curve ($C_{\rm v}$) shows a cusp.
Such cusps in specific-heat versus temperature plots have been observed recently in experiments for materials with two-step phase transitions\cite{CrCl3}.
In \FeCl, the difference in the phase transition is very prominent.
Lowering the temperature below this phase transition temperature, the magnetization of \FeCl~tends towards zero, which implies that this is an anti-ferromagnetic phase transition.
Because this state evolves from the intralayer ferromagnetic phase, there is smearing in the specific-heat data, hence, instead of a number for the N\'eel temperature, we consider the N\'eel temperature to be $47\pm 8$ K, which differs only by a factor of two with the previously reported experimental value of $24$ K~\cite{FeCl2_exp}.

To further understand the magnetic behavior of \FeCl, we show the magnetic order in \FeCl~at various temperatures in Fig.~\ref{f:FeCl2_GS}.
At a temperature of $386$ K, we observe a paramagnetic phase with randomly aligned magnetic moments of Fe atoms.
The first phase transition in \FeCl~occurs below 250 K.
At a temperature of $160$ K, we see that the layers of \FeCl~are aligned ferromagnetically, confirming that the high-temperature phase transition is intralayer ferromagnetic in nature.
In the out-of-plane direction, the magnetic orientation is paramagnetic, and each of the layers of \FeCl~start behaving like a single unit, where all the magnetic moments of each layer change their orientation simultaneously.
%The paramagnetic orientation in the out-of-plane direction causes the maximum magnetic moment to reach a value of $0.9\,\mu_{\rm B}$ per Fe atom instead of reaching its ideal maximum value of $3.5\,\mu_{\rm B}$.
At a temperature of $70$ K, we see a transition from a paramagnetic to an anti-ferromagnetic orientation in the out-of-plane direction, which persists until $T=50$ K.
At $T=38$ K, we see an anti-ferromagnetic alignment in the out-of-plane direction, but due to weak geometric and magnetic anisotropy of \FeCl, the anti-ferromagnetic alignment is not perfect yet.
However, we see that the phase transition occurs between $70-38$ K.
Finally, at a temperature of $6$ K the anti-ferromagnetic ground state is reached, showing an ideal atomic N\'eel pattern, with a saturation magnetization as low as $0.1\,\mu_{\rm B}$ per Fe atom.
The interlayer anti-ferromagnetic phase of \FeCl~evolves from the intralayer ferromagnetic phase and is similar in nature when compared to the experimental observation of the two-step phase transition in $\rm CrCl_3$~\cite{CrCl3}.

\begin{figure}[ht]
   \centering
  \includegraphics[width=\columnwidth]{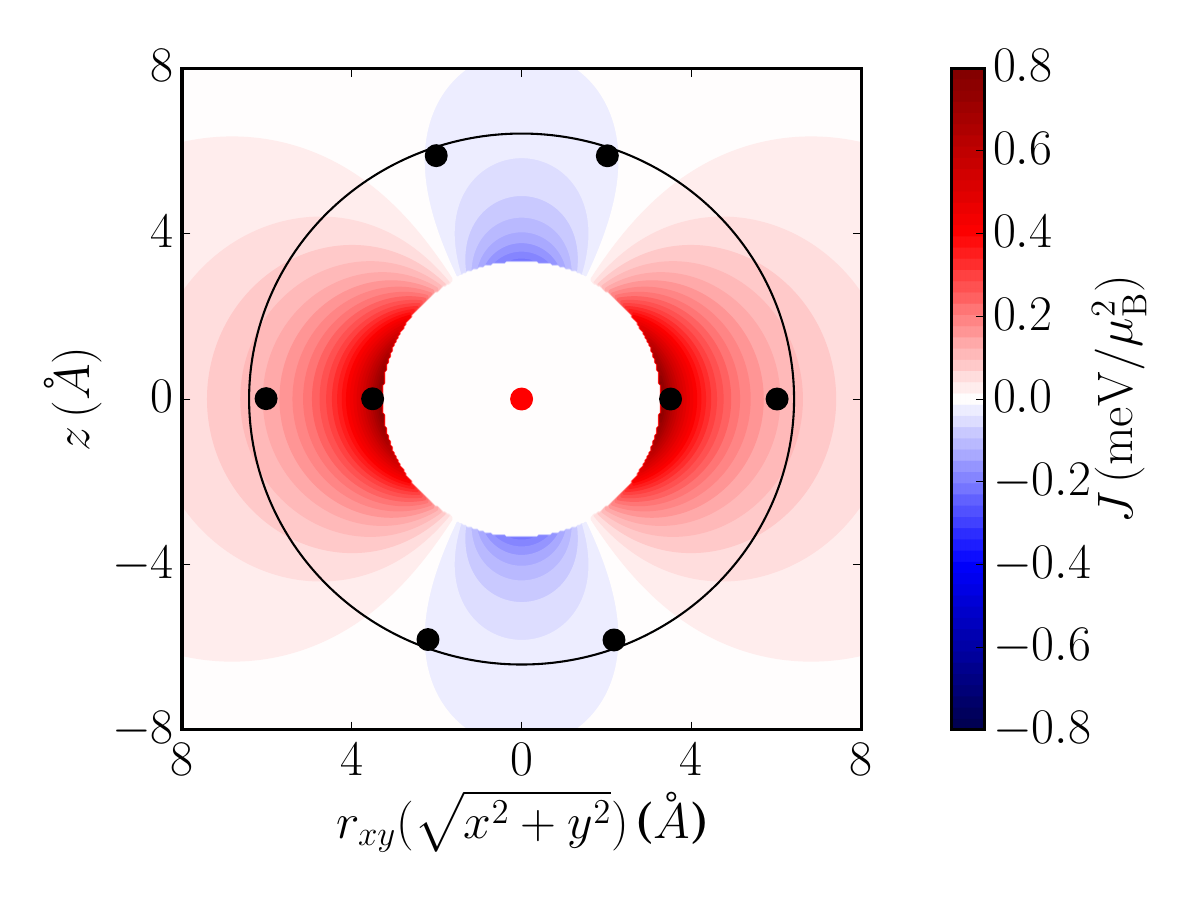}
   \caption{The $J$ parameters for the \FeCl~bulk with magnetic easy axis in the out-of-plane direction ($J^{zz}(r,\theta)$). The circle shows the range of the interaction ($r_{\rm c}=6.22$ \AA) with respect to the atom at the centre (in red). The atoms inside the interaction circle are the ones considered while obtaining the $J$ parameters.}
  \label{f:FeCl2_ex}
\end{figure}

Figure~\ref{f:FeCl2_ex} shows the exchange function ($J^{zz}(r,\theta)$) with magnetic dipoles ($\mathbf{S}$) oriented along its magnetic easy axis ($z$) for \FeCl.
We find that the in-plane exchange interaction $J$ in \FeCl~is very strong, with a nearest-neighbor interaction strength of $0.64\, \rm meV/\mu_B^2$, compared to the nearest-neighbor interaction strength of \CrI which is, $0.24\, \rm meV/\mu_B^2$.
The out-of-plane interaction in \FeCl~is weak and anti-ferromagnetic with $J_{\mathrm{out}}=-1.8\times10^{-2}\, \rm meV/\mu_B^2$.
The strong in-plane interaction is consistent with the previous DFT based studies of monolayer \FeCl~\cite{FeCl2}.
Also, the out-of-plane anti-ferromagnetic orientation is in accordance with the previous experimental reports~\cite{FeCl2_exp}.
Thanks to the strong in-plane and weak out-of-plane interaction, the double phase transition of \FeCl~shown in Fig.~\ref{f:Curie_FeCl} is very prominent.

\section{Computational model}{\label{s:method2}}

In this section, we provide the details of our computational model, detailing how we construct the Heisenberg Hamiltonian to obtain the $J$ parameters.
We provide the details regarding the Monte-Carlo simulations we employ to simulate the phase transition using the parameterized Heisenberg Hamiltonian.

We calculate the critical temperature from first-principles DFT calculations without any input from experiments.
Our method treats both magnetic and geometric anisotropy simultaneously, which is not possible using the methods which use direct energy difference~\cite{approach_1, CrI_bad, CrGeTe_bad}.
Moreover, as discussed in Ref.~\cite{S_problem}, the daunting task of modeling helimagnetic configurations, \ie comparing energies of magnetic configurations with a different magnetic axis, is included in our method.
%Although our approach is computationally more expensive than the simple mean-field estimates, it is much cheaper compared to methods like dynamical mean-field theory (DMFT)~\cite{DMFT}. 
%It is significantly more accurate, yielding very good results for magnetic materials having FM or anti-ferromagnetic ground state, as shown in section~\ref{s:Curie_temp}.

\subsection{Obtaining the exchange parameters}

\begin{figure}[th]
	\centering
   \includegraphics[width=0.9\columnwidth]{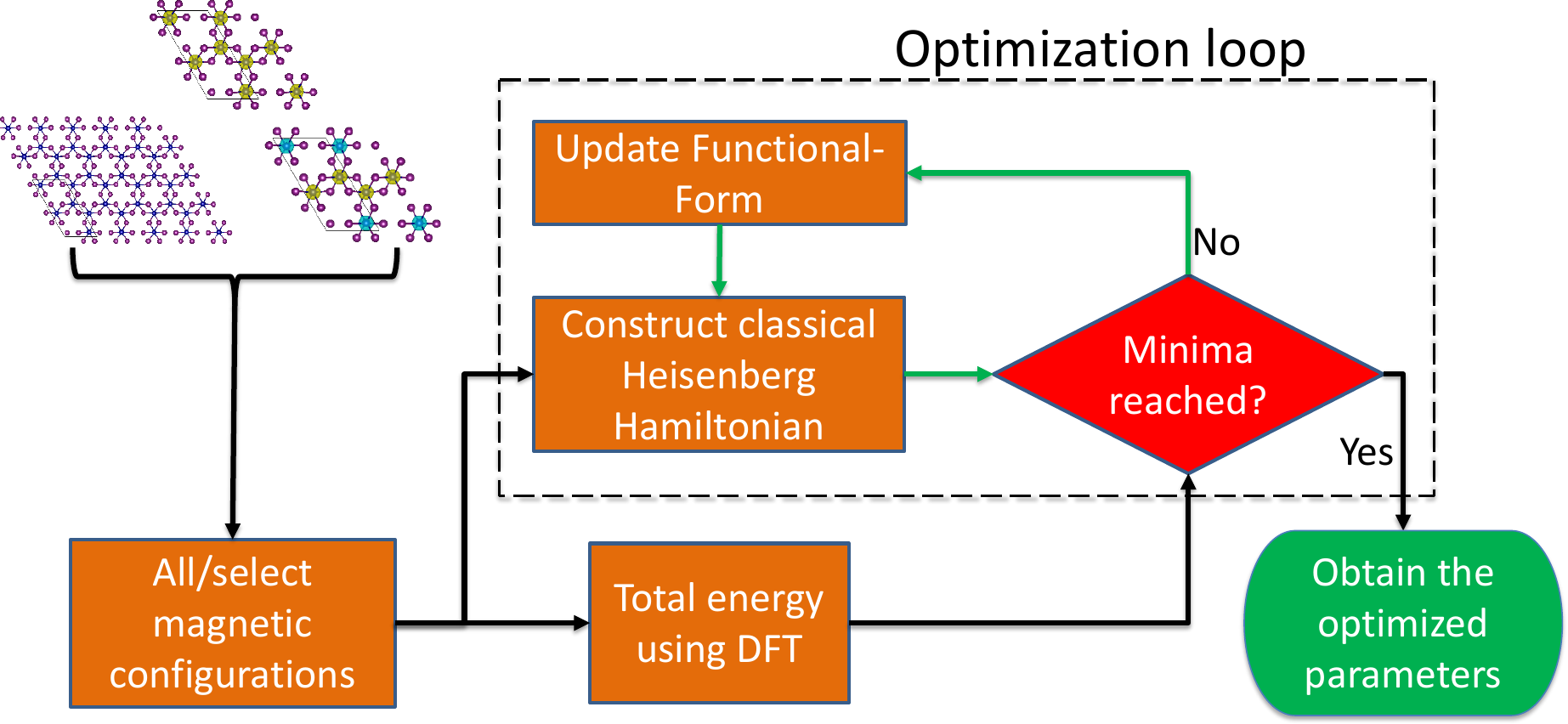}
    \caption{The various blocks of our computational model.}
	\label{f:flowchart}
\end{figure}

Figure~\ref{f:flowchart} shows the flowchart of our computational procedure to obtain the $J$ parameters.
The crystal structure is the only overall input.
Details of each of the computational block are provided in the subsequent sections.

\subsubsection{Total energy calculation for various magnetic configurations}{\label{s:supercells}

\begin{figure}[th]
  \centering
  \subfigure[]{\includegraphics[width=.45\columnwidth]{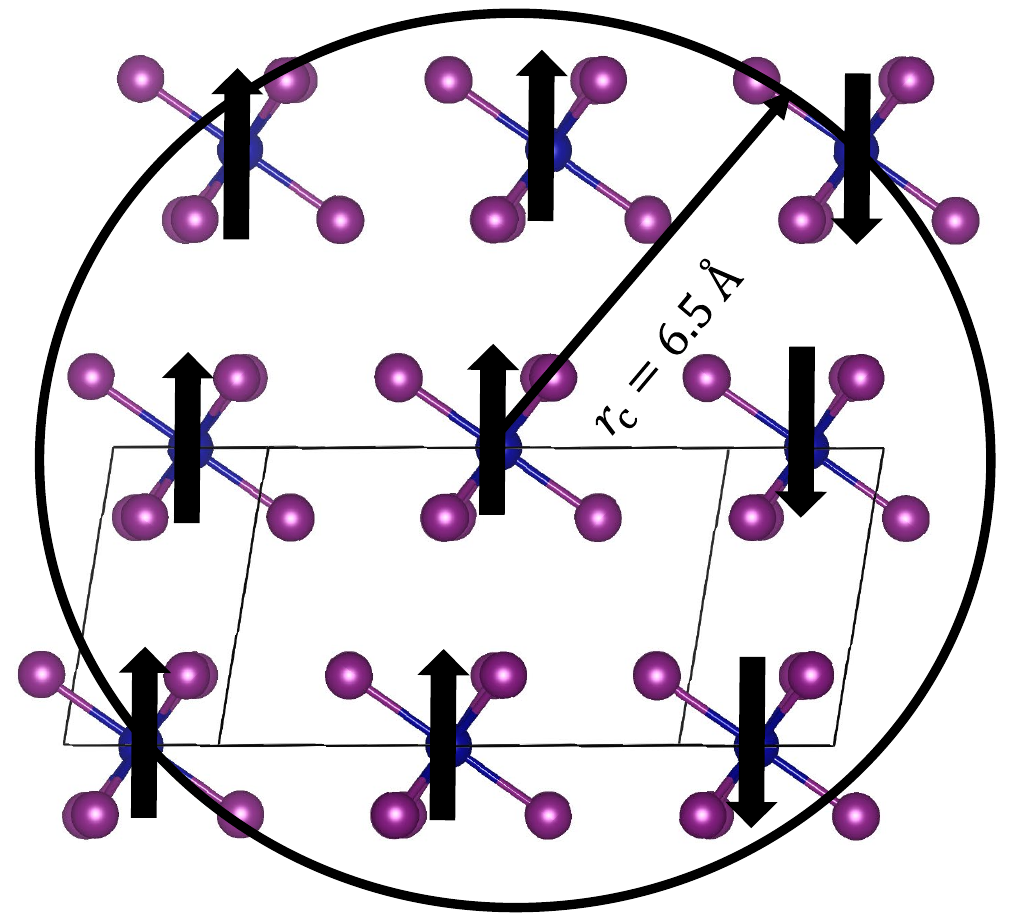}}
  %\vspace{\floatsep}
  \subfigure[]{\includegraphics[width=.45\columnwidth]{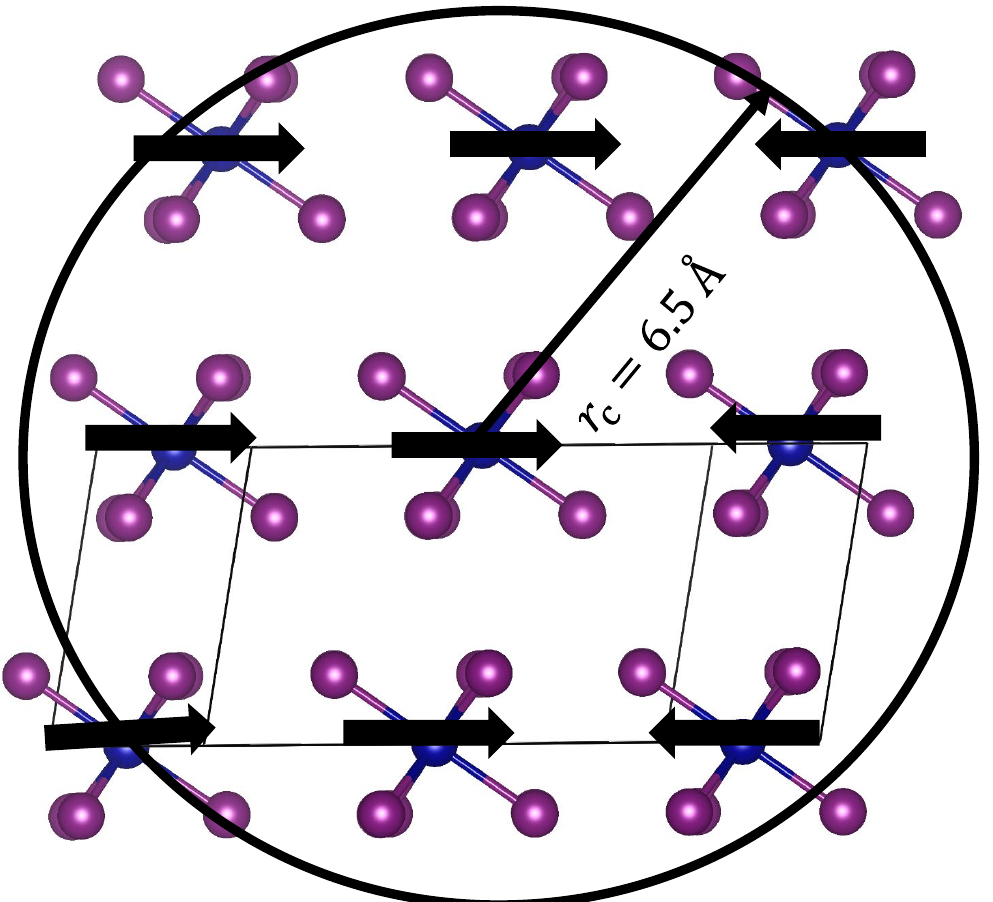}}
\vspace{\floatsep}
%  \subfigure[]{\includegraphics[width=.45\columnwidth]{CHGCAR_N_spn4.pdf}}
  \caption{Illustration of a ferrimagnetic configuration of \CrI with (a) out-of-plane magnetic easy axis, and (b) in-plane magnetic easy axis. The circle shows the range up to which the exchange interaction is considered.} 
  \label{f:scheme_2} 
\end{figure} 
For DFT, we build supercells of size $N_{\mathrm{x}}\times N_{\mathrm{y}}\times N_{\mathrm{z}}$ by repeating the primitive unit-cell of the magnetic material $N_{\mathrm{x}}/N_{\mathrm{y}}/N_{\mathrm{z}}$ times in the $x/y/z$ direction.
The size of the supercells ($N_{\mathrm{x}}/N_{\mathrm{y}}/N_{\mathrm{z}}$) is set by the range of the interaction, quantified by the radius of interaction ($r_{\mathrm{c}}$).
Within the radius of interaction $r_{\mathrm{c}}$, the total energy of all the possible magnetic configurations, \ie anti-ferromagnetic, ferromagnetic, and ferrimagnetic configurations is calculated using DFT.
For example, a supercell of \CrI (\C) is shown in Fig.~\ref{f:scheme_2} with a sphere of interaction $r_{\rm c} = 6.5\,\AA$, and with a ferrimagnetic configuration.

To account for the magnetic anisotropy, we use magnetic configurations with the magnetic axis of all the magnetic atoms oriented in different planes (out-of-plane along $z$ $(001)$, in-plane along $y$ $(010)$, mixed plane along $yz$ $(011)$), and calculate their total energy using non-collinear DFT calculations.
For example, Fig.~\ref{f:scheme_2} (a) shows a magnetic configuration with the magnetic axis in the out-of-plane $(001)$ direction while (b) shows the same magnetic configuration but with magnetic axis in the in-plane $(110)$ direction.

\subsubsection{Functional-form for the exchange parameters}{\label{s:func_form}}

As mentioned earlier in section~\ref{s:method}, we use a functional form to describe the $J$ parameters.
Using a functional form $J(r)$ with a physical exponential decay at the long range, instead of discrete $J_ij$ parameters between each pair $ij$ reduces the number of parameters to be determined to describe accurately long-range interactions.

For brevity, we ignore the superscripts of the exchange interactions, \eg~$J^{xx}(r,\theta)$, in this subsection. 
We introduce a rotationally invariant exchange function with in-plane isotropy~\cite{sample7,sample8},

\begin{multline}
    J(r_{ij},\theta_{ij})  = \frac{\left(c_\parallel\cos(\theta_{ij})^2+c_\perp\sin(\theta_{ij})^2\right)}{r_{ij}^3}\\
     \exp\left(-r_{ij}\sqrt{\left(\frac{\cos(\theta_{ij})^2}{\lambda_\parallel^2}+\frac{\sin(\theta_{ij})^2}{\lambda_\perp^2}\right)}\right)u(r_c-r_{ij}) 
    \label{e:J}
\end{multline}
Here, $u(r)$ is the Heavyside function limiting the range of the interaction to the radius $r_{\rm c}$.
For 2D materials that are isotropic in the layer, there are two sets of parameters, $c_\perp,\lambda_\perp$ for the out-of-plane magnetic interaction, and $c_\parallel,\lambda_\parallel$ for the in-plane magnetic interaction.
$r_{ij}$ is the distance between the magnetic ions with $r_{ij}=|r_i-r_j|$.
$\theta_{ij}$ is the azimuthal (out-of-plane) angle between the magnetic atom $i$ and $j$ (refer to Fig.~\ref{f:illustration_xyz}).

The functional-form in Eq.~(\ref{e:J}) has a long-range exponential screening ($\propto \exp(-r_{ij}/\lambda)$) and a short-range cubic screening ($\propto 1/r_{ij}^3$).
These asymptotic behaviors match both the short-range magnetic interaction (cubic) and the long-range exchange interaction (exponential)~\cite{sample8}.

Note that for any future calculations on 2D single layers, the derivation of the functional-form is trivial with $\theta_{ij}=0$ leading to,
\begin{equation}
    J(r_{ij},\theta_{ij})=\frac{c_\parallel}{r_{ij}^3}\exp\left(-\frac{r_{ij}}{\lambda_\parallel}\right)u(r_c-r_{ij}) 
    \label{e:J}
\end{equation}
Here, only the parameters $c_\parallel,\lambda_\parallel$ are optimized.

\subsubsection{Building the Heisenberg Hamiltonian}

To obtain the parameters of the exchange function, we build the classical Heisenberg Hamiltonian for various magnetic configurations and fit the calculated energies for each of the magnetic configurations to the energy obtained from the DFT calculations.
We parameterize the Heisenberg Hamiltonian using continuous $J(r_{ij},\theta_{ij})$ functions, and optimize the parameters of $J(r_{ij},\theta_{ij})$ to fit the energies obtained from the DFT calculations (more details on optimization is provided in section.~\ref{s:optimization}). 

We implement the full $J$ tensor in our computational model. 
However, to explain our method here, we omit the off-diagonal elements of tensor $J(r_{ij},\theta_{ij})$ of Eq.~(\ref{e:tensor}) and expand the Heisenberg Hamiltonian as, 

\begin{multline}
   H
    = -\sum_{i\neq j} S_{i}^xJ^{xx}(r_{ij},\theta_{ij})S_{j}^x-\sum_{i\neq j} S_{i}^yJ^{yy}(r_{ij},\theta_{ij})S_{j}^y\\
-\sum_{i\neq j} S_{i}^zJ^{zz}(r_{ij},\theta_{ij})S_{j}^z-\sum_{i} S_{i}^zDS^z_{i}.
    \label{e:hamiltonian_anis_2}
\end{multline}

We convert the magnetic moment components in their polar form using, 
\begin{subequations}
\begin{align}
S^x=S\cos(\Phi)\sin(\Theta), \\
S^y=S\sin(\Phi)\sin(\Theta), \\
S^z=S\cos(\Theta)\\
S=\sqrt{(S^x)^2+(S^y)^2+(S^z)^2}, \\ 
\Phi= \tan^{-1}(S_y/S_x), \\  
\Theta=\tan^{-1}(S_z/\sqrt{S_x^2+S_y^2})
\end{align}
\end{subequations}
The resulting Hamiltonian with in-plane isotropy ($J^{xx}(r_{ij},\theta_{ij})=J^{yy}(r_{ij},\theta_{ij})=J^{\parallel}(r_{ij},\theta_{ij})$) reads,
\begin{multline}
   H   
    = -\sum_{i\neq j} J^{\parallel}(r_{ij},\theta_{ij})S_{i}S_{j}(\cos(\Phi_i-\Phi_j)\sin(\Theta_i)\sin(\Theta_j))\\
     -\sum_{i\neq j}J^{zz}(r_{ij},\theta_{ij}) S_{i}S_{j} \cos(\Theta_i)\cos(\Theta_j)\\ 
     -\sum_{i}D(S_{i}\cos(\Theta_i))^2.
    \label{e:hamiltonian_anis3}
\end{multline}

\begin{figure}[ht]
  \centering
  \includegraphics[width=.9\columnwidth]{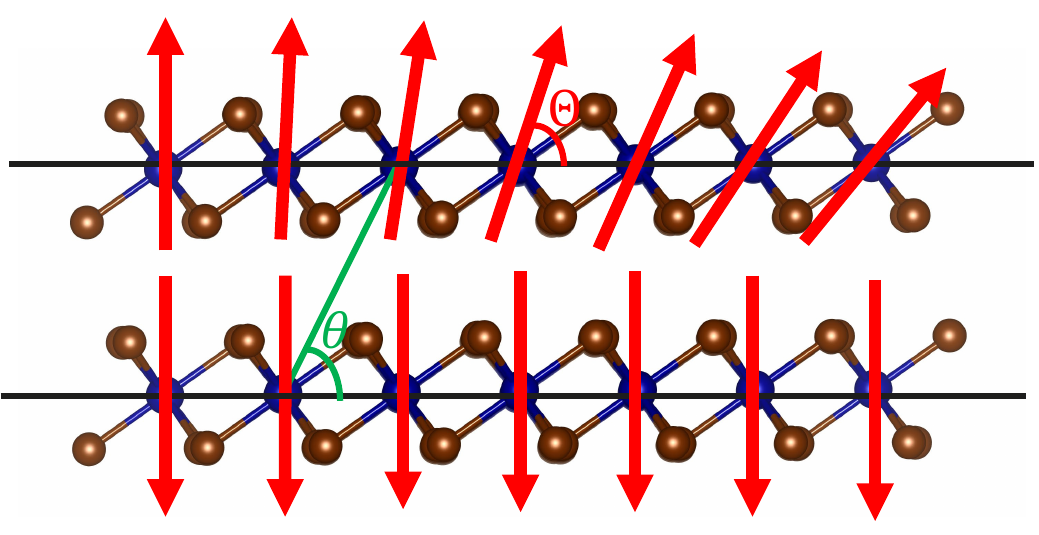}
   \caption{Illustration of the geometric ($\theta$) and the magnetic azhimuthal angle ($\Theta$).}
 \label{f:illustration_xyz}
\end{figure}

To illustrate the difference between $\theta$ and $\Theta$, Fig.~\ref{f:illustration_xyz} shows the geometric angle $\theta$ (green) between the Cr atoms of two layers in \CrI whereas, angle $\Theta$ (in red) shows the angle of the magnetic moment of a particular Cr atom in $\rm CrI_3$.
We have dropped the indices of $\theta$ and $\Theta$ in Fig.~\ref{f:illustration_xyz} for brevity.
%Here, $\theta$ is the geometric angle between the atoms at two different layers and $\Theta$ is the magnetic angle of a magnetic atom.
%An illustration showing the difference between the $\theta$ and $\Theta$ is depicted in Fig.~\ref{f:illustration2}(d). 

The Hamiltonian in Eq.~(\ref{e:hamiltonian_anis3}) is now written for each of the magnetic configurations with an index $l$ as a function of exchange functions, $J^{\parallel}(r_{ij},\theta_{ij})$ and $J^{zz}(r_{ij},\theta_{ij})$, and the onsite anisotropy ($D$),
\begin{multline}
   H_l(J^{\parallel},J^{zz},D)   
    =- \sum_{i\neq j}S^l_{i}S^l_{j}\{ J^{\parallel}(r_{ij},\theta_{ij})(\cos(\Phi^l_i-\Phi^l_j)\\
     \sin(\Theta^l_i)\sin(\Theta^l_j))+J^{zz}(r_{ij},\theta_{ij})\cos(\Theta^l_i)\cos(\Theta^l_j)\}\\ 
     -\sum_{i}(S^l_{i}\cos(\Theta^l_i))^2(D).
    \label{e:hamiltonian_anis3}
\end{multline}
We build the Heisenberg Hamiltonian for all the magnetic configurations for which we calculate the total energy ($E_l$) and the magnetizations $\mathbf{S}$ using DFT, as discussed in section~\ref{s:supercells}.

In the case of magnetic structures with helimagnetic configurations (off-diagonal elements of $J_{ij}$), mixed terms will be added to Eq.~(\ref{e:hamiltonian_anis3}), \eg for a term like $S^x_iJ^{xy}(r_{ij},\theta_{ij})S^y_j$ we get, $\sum_{i\neq j}S_iS_jJ^{xy}(r_{ij},\theta_{ij})(\cos(\Phi^l_i)\sin(\Phi^l_j)\sin(\Theta^l_i)\sin(\Theta^l_j))$.

\subsubsection{Optimization of the exchange interaction parameters}{\label{s:optimization}}

We fit the Heisenberg Hamiltonian, $H_{l}(J^{\parallel},J^{zz},D)$, to the total energy ($E_l$) calculated using the DFT calculations for the same magnetic configurations.
We fit using a least-squares objective function,

\begin{equation}
    \mathrm{O}
    = \sqrt{\sum_l|E_{l}-E_{\mathrm{NM}}-H_{l}(J^{\parallel},J^{zz},D)- H_{l}^{\mathrm{self}}|^2},
    \label{e:objectivefunction}
\end{equation}
and the differential evolution technique~\cite{diff_evol_1,diff_evol_2} to obtain the parameters of the exchange functions $J^{\parallel/zz}(r_{ij},\theta_{ij})$, and the onsite anisotropy $D$.

In Eq.~(\ref{e:objectivefunction}), $E_{\mathrm{NM}}$ is the total energy of the non-magnetic configuration obtained using DFT.
$H_{l}^{\mathrm{self}}$ is the self-energy interaction term for the $l^{\rm th}$ magnetic configuration,
\begin{equation}
H_{l}^{\mathrm{self}} = \sum_{i}t(|\mathbf{S}^l_{{i}}|)^2.
\end{equation}
Here, $t$ is the self-interaction for the $l^{\rm th}$ magnetic configuration, and $S^l_{{i}}$ is the magnetic moment for the $l^{\rm th}$ magnetic configuration and the $i^{\rm th}$ magnetic atom.
We find that the contribution of self-interaction is very-small in terms of the energy difference ($\Delta E$) calculated between various magnetic configurations.

%\fix{The optimization procedure is very general and instead of using a functional form, discrete $J$ parameters can also be evaluated. 
%However, using a functional form reduces the number of parameters to be evaluated to include long-range interactions.}

\subsection{Obtaining critical temperatures}

We calculate the critical temperature of a material using Monte-Carlo simulations.
The Monte Carlo simulations are performed on supercells built from the original crystal.
The supercells we employ here are much larger than those we use in the DFT calculations to determine the parameters of the exchange functional.
Every atom in the Monte Carlo simulation domain has a spin configuration which is flipped based on a localized Heisenberg Hamiltonian.
Following we describe how the localized Heisenberg Hamiltonian is constructed and how the Monte Carlo algorithm proceeds.

\subsubsection{Mapping $J$ to Heisenberg Hamiltonian}

The localized Heisenberg Hamiltonian is,
\begin{equation}
    H
    = -M^2 \sum_{i\neq j}\hat{\sigma_{i}}J(r_{ij},\theta_{ij})\hat{\sigma_{j}}-M^2 \sum_{i}{\sigma^z_{i}}D{\sigma^z_{j}}
    \label{e:classical_hamiltonian2}
\end{equation}
where $J(r_{ij},\theta_{ij})$ is the exchange interaction tensor between the $i^{\rm th}$ and the $j^{\rm th}$ magnetic atom.
$\hat{\sigma}_{i/j}={\sigma}^x_{i/j}\mathbf{x}+{\sigma}^y_{i/j}\mathbf{y}+{\sigma}^z_{i/j}\mathbf{z}$, is the spin-polarization vector of the $i^{\mathrm{th}}/j^{\mathrm{th}}$ magnetic atom.
$D$ is the onsite anisotropy.

All information from the DFT calculations is contained in $J(r,\theta)$ and $D$ except for the value of the magnetic moment $M$.
Since every atom in the different simulated magnetic configuration gives rise to a slightly different magnetic moment.
We calculate the average magnetization as,
\begin{equation}
M=\frac{1}{N_\mathrm{c}}\sum_l\frac{1}{N}\sum_jM_{\mathrm{DFT}}^{l,j}
\label{e:mag_mom}
\end{equation}
Here, $M_{\mathrm{DFT}}^{l,j}$ is the average of the magnetic moment of the $j^\mathrm{th}$ magnetic atom of $l^\mathrm{th}$ magnetic configuration obtained from DFT, $M_{\mathrm{DFT}}^{l,j}=|\mathbf{S}^{l}_j|$ with ${\mathbf{S}}=S^x\mathbf{x}+S^y\mathbf{y}+S^z\mathbf{z}$.
$N_\mathrm{c}$ and $N$ are the total number of the magnetic configurations simulated and the magnetic atoms, respectively.

\subsubsection{Monte-Carlo algorithm}\label{s:monte-carlo}

We use the Monte-Carlo method to simulate the magnetic phase-change using the Metropolis algorithm~\cite{sample9}.
For the Monte-Carlo calculations of bulk materials, we use periodic boundary conditions. 

The Monte-Carlo algorithm goes as follows:
\begin{itemize}
%\item[$\bullet$]{The supercell of the crystal is initiated with a random magnetic orientation for each of the magnetic ion.} 
\item[$\bullet$]For a particular temperature, we perform $N_{\rm MC}$ Monte-Carlo (MC) steps.
\item[$\bullet$]For each MC step, $N_{\rm atom}$ spin-flip steps are performed where, for each spin-flip step, an atom ($i$) of the chosen supercell is selected at random and its magnetic polarization ($\sigma_i$) is rotated randomly with a uniform probability. The polarization vector can be rotated by multiplying it with a rotator vector: $\sin(\Phi)\sqrt{1-u^2}\mathbf{x}+\cos(\Phi)\sqrt{1-u^2}\mathbf{y}+u\mathbf{z}$, with $u \in [-1,1]$, and in-plane angle, $\Phi \in [0,2\pi]$, which yields a uniform sampling of all angles.
\item[$\bullet$]If the resulting magnetic moment results in a lowering of the total energy, the resulting magnetic moment replaces the previous one. Otherwise, the new magnetic moment is replaces the previous one with a probability $\exp\left(\frac{\Delta E}{k_{\mathrm{B}}T}\right)$. Where, $\Delta E$ is the difference between the total energy of the material system before and after the magnetic moment of the chosen atom was rotated, calculated using Eq.~(\ref{e:classical_hamiltonian2}).
\end{itemize}

The number of the spin-flip steps ($N_{\rm atom}$) is equal to the number of atoms in the supercell, whereas, the number of MC steps ($N_{\rm MC}$) is chosen based on the size of the structure as well as the exchange parameters ($J$).
For each temperature, we start with the magnetic configuration of the previous temperature and perform $N_{\rm{eq}}$ MC steps, which are not used for taking the average but help in equilibrating the system.

From the Monte-Carlo calculations, we obtain the average magnetization $\langle M\rangle$ as a function of temperature ($T$) as,

\begin{equation}
\langle M(T)\rangle=\frac{1}{N_{\mathrm{MC}}N_{\mathrm{atoms}}}\sum_{i}^{N_{\mathrm{MC}}}\sum_{j}^{N_{\mathrm{atoms}}} M_{ij}(T).
\label{e:magnetization}
\end{equation}
We also calculate the absolute average magnetization by taking $|M_{ij}(T)|$ instead of $M_{ij}(T)$ in Eq.~(\ref{e:magnetization}). 
The susceptibility is calculated as the second moment of the magnetization $M$,
\begin{equation}
    \chi (T)
    = \frac{1}{k_{\mathrm{B}}T}(\langle{M(T)^2}\rangle-\langle{M(T)}\rangle^2).
    \label{e:susceptibility}
\end{equation}
The temperature at which the susceptibility ($\chi (T)$) peaks is the critical temperature (Curie/N\'eel) for a magnetic phase transition (ferromagnetic/anti-ferromagnetic).
Hence, $T_{\mathrm{C}}=\mathrm{argmax}(\chi (T))$.

We also obtain the total energy $E(T)$ as a function of temperature from the MC simulations using which we calculate the specific-heat as,
\begin{equation}
    C_{\mathrm{v}} (T)
    = \frac{1}{k_{\mathrm{B}}T^2}(\langle{E(T)^2}\rangle-\langle{E(T)}\rangle^2).
    \label{e:susceptibility}
\end{equation}

In material systems with complicated phase transitions, critical temperatures are the temperatures at which the specific-heat peaks.
Multiple peaks of the specific-heat suggest multiple phase transitions.

Due to statistical noise near criticality, we use the Savitzky-Golay (savgol) filter to cancel the statistical noise in the MC magnetic moment and specific heat as a function of temperature.
We use a window range of 51 (total window 180) and a polynomial of order 5 for the savgol filter.

To calculate the Curie temperature of ferromagnets $\rm CrI_3$, \CrBr~and $\rm CrGeTe_3$ using MC simulations, we use supercells of size $6\times6\times2$, which results in 216 $\rm Cr$ atoms for $\rm CrI_3$,~\CrBr,~and 300 $\rm Cr$ atoms for \CrGeTe.
For taking the average in the Monte-Carlo simulations, we use 3000 steps ($N_{\rm MC}=3000$).
Also, we use 3000 steps for equilibration at the start of each temperature cycle.

To calculate the Ne\'el temperature of \FeCl~using MC simulations, we use a $5\times5\times2$ supercell of \FeCl~(216 $\rm Fe$ atoms).
We use 1000 equilibration steps for equilibrating the magnetic structure at the start of every temperature cycle and 1000 MC steps for averaging the observables.

We find that the chosen number of equilibration and MC steps are sufficient for reaching thermal equilibrium and calculating average quantities, respectively.
More details on equilibration and convergence are provided in the supplementary information.

\subsection{DFT calculations}

All the first-principles DFT calculations reported in this work were performed using the Vienna ab-initio simulation package (VASP)~\cite{sample3,sample5}.
The ground state self-consistent field SCF calculations were performed using the projector-augmented wave (PAW) method~\cite{sample3} within the generalized gradient approximation, as proposed by Perdew-Burke-Ernzerhof (PBE)~\cite{sample4}.
To take into account the Van der Waals interaction in layered magnetic materials, we use the DFT$+$D3 method \cite{vdw1,vdw2}.
For the ionic relaxation, we relax the crystal structure until the force on each of the ions is less than 5 meV/\AA.
For the SCF convergence, we use an energy threshold of $10^{-4}$ eV.
We use a Monkhorst-Pack $k$-point sampling scheme~\cite{MP} of $5\times5\times5$.
The plane-wave energy-cutoff scheme was set to 400 eV for all the materials reported in this work.

To calculate the $J$ parameters of \R~$\rm CrI_3$, \CrBr~and $\rm CrGeTe_3$, we use supercells (periodic repitition of the primitive unit-cell) of size $2\times 1\times 1$, $1\times 2\times 1$ and the unit-cell (shown in Fig.~\ref{f:CrI_R3}).
We use non-collinear (NCL) DFT calculations with spin-orbit coupling for calculating the total energy of the magnetic configurations.
The magnetic axis we choose for the NCL calculations are $(001),\,(010)$ and $(0\frac{1}{\sqrt{2}}\frac{\sqrt{3}}{2})$.
For each of the chosen axes, we simulate all the magnetic configurations within a cut-off radius of 6.81 \r{A}.

For \C~$\rm CrI_3$, we use supercells of size $2\times1\times1$, $1\times2\times1$ and $1\times1\times2$, with a cut-off radius of 7.1 \r{A} for evaluating the $J$ parameters.
We use the same set of magnetic axis for the NCL calculations as for the \R~$\rm CrI$.

To calculate the $J$ parameters of \FeCl, we use a supercell of size $2\times2\times1$.
We orient the magnetic axis in the $(001)$, $(010)$, and $(100)$ direction for the total energy calculations.
The lattice structure of \FeCl~is isotropic in the in-plane direction with $(001)$ as the direction of the magnetic easy axis.

\section{Conclusion}

We have calculated the Curie temperature of ferromagnets \CrI, \CrBr~and \CrGeTe~to be 72 K, 49 K, and 103 K, whose experimentally measured values are, 61 K, 37 K and $66$ K, respectively.
We have calculated the N\'eel temperature of \FeCl~to be $47\pm8$ K. %which is of the same order as its experimentally measured value of $24$ K.
Moreover, we have matched the near-criticality behavior of the Cr compounds by estimating their critical exponent $\beta=0.312$ for $\rm CrI_3$, while experimentally $\beta=0.325$.

We have shown that the out-of-plane interaction in \R~stacked \CrI is stronger than the \C~stacked $\rm CrI_3$.
The strong out-of-plane interaction in \R~\CrI and the weak out-of-plane interaction in \C~\CrI is in line with experimental observations.
We have shown that the strong out-of-plane interaction in \R~\CrI results in a competition for the magnetic easy axis orientation.
However, the magnetic easy axis orients itself in the out-of-plane direction thanks to the increased in-plane interactions, lowering the total energy.

In \FeCl, we have shown that the magnetic interaction is strongly ferromagnetic in-plane and weakly anti-ferromagnetic out-of-plane.
We have shown that \FeCl~undergoes two phase transitions, a high-temperature phase transition with in-plane FM order and out-of-plane paramagnetic order, and a low-temperature phase transition with out-of-plane anti-ferromagnetic order and in-plane FM order.

We have presented a method to calculate the magnetic exchange ($J$) parameters and the critical temperature of magnetic materials from first principles.
Our method is very general and can be applied for both monolayer and multi-layered magnetic materials, and yields much better results, compared to the previous theoretical works for the layered ferromagnetic compounds $\rm CrI_3$, \CrBr, and \CrGeTe.

\section{Acknowledgements}
The project or effort depicted was or is sponsored by the Department of Defense, Defense Threat Reduction Agency.
The content of the information does not necessarily reflect the position or the policy of the federal government, and no official endorsement should be inferred.

This work was supported by imec's Industrial Affiliation Program.

\section*{References}
\bibliography{bib}

\end{document}